\documentclass[12pt]{article}
\usepackage{graphicx}

\def\hybrid{\topmargin 0pt      \oddsidemargin 0pt
        \headheight 0pt \headsep 0pt
       \voffset-1cm
        \textwidth 6.25in       
       \textheight 9.5in       
        \marginparwidth 0.0in
        \parskip 5pt plus 1pt   \jot = 1.5ex}
\catcode`\@=11
\def\marginnote#1{}

\newcount\hour
\newcount\minute
\newtoks\amorpm
\hour=\time\divide\hour by60
\minute=\time{\multiply\hour by60 \global\advance\minute by-\hour}
\edef\standardtime{{\ifnum\hour<12 \global\amorpm={am}%
        \else\global\amorpm={pm}\advance\hour by-12 \fi
        \ifnum\hour=0 \hour=12 \fi
        \number\hour:\ifnum\minute<10 0\fi\number\minute\the\amorpm}}
\edef\militarytime{\number\hour:\ifnum\minute<10 0\fi\number\minute}

\def\draftlabel#1{{\@bsphack\if@filesw {\let\thepage\relax
   \xdef\@gtempa{\write\@auxout{\string
      \newlabel{#1}{{\@currentlabel}{\thepage}}}}}\@gtempa
   \if@nobreak \ifvmode\nobreak\fi\fi\fi\@esphack}
        \gdef\@eqnlabel{#1}}
\def\@eqnlabel{}
\def\@vacuum{}
\def\draftmarginnote#1{\marginpar{\raggedright\scriptsize\tt#1}}

\def\draftlabel#1{{\@bsphack\if@filesw {\let\thepage\relax
   \xdef\@gtempa{\write\@auxout{\string
      \newlabel{#1}{{\@currentlabel}{\thepage}}}}}\@gtempa
   \if@nobreak \ifvmode\nobreak\fi\fi\fi\@esphack}
        \gdef\@eqnlabel{#1}}
\def\@eqnlabel{}
\def\@vacuum{}
\def\draftmarginnote#1{\marginpar{\raggedright\scriptsize\tt#1}}

\def\draft{\oddsidemargin -.5truein
        \def\@oddfoot{\sl preliminary draft \hfil
        \rm\thepage\hfil\sl\today\quad\militarytime}
        \let\@evenfoot\@oddfoot \overfullrule 3pt
        \let\label=\draftlabel
        \let\marginnote=\draftmarginnote
   \def\@eqnnum{(\theequation)\rlap{\kern\marginparsep\tt\@eqnlabel}%
\global\let\@eqnlabel\@vacuum}  }

\def\numberbysection{\@addtoreset{equation}{section}
        \def\theequation{\thesection.\arabic{equation}}}

\def\underline#1{\relax\ifmmode\@@underline#1\else
        $\@@underline{\hbox{#1}}$\relax\fi}

\def\titlepage{\@restonecolfalse\if@twocolumn\@restonecoltrue\onecolumn
     \else \newpage \fi \thispagestyle{empty}\c@page\z@
        \def\thefootnote{\fnsymbol{footnote}} }

\def\endtitlepage{\if@restonecol\twocolumn \else  \fi
        \def\thefootnote{\arabic{footnote}}
        \setcounter{footnote}{0}}  
\relax

\numberbysection
\hybrid

\newfont{\Bbb}{msbm10 scaled 1\@ptsize00}
\newfont{\Bbbb}{msbm7 scaled 1\@ptsize00}

\newcommand{\DDD}{\raise-1pt\hbox{$\mbox{\Bbbb D}$}}

\newcommand{\UUU}{\raise-1pt\hbox{$\mbox{\Bbbb U}$}}

\newcommand{\ZZ}{\mbox{\Bbb Z}}
\newcommand{\z}{\raise-1pt\hbox{$\mbox{\Bbbb Z}$}}

\newcommand{\sss}{\raise-1pt\hbox{$\mbox{\Bbbb S}$}}

\def\beq{\begin{equation}}
\def\eeq{\end{equation}}
\def\p{\partial}

\newtheorem{lemma-definition}{Lemma-Definition}[section]

\begin{document}

\begin{titlepage}

\title{Dispersionless limit of the B-Toda hierarchy}

\author{A. Zabrodin\thanks{Skolkovo Institute of Science and
Technology, 143026, Moscow, Russia and
National Research University Higher School of
Economics,
20 Myasnitskaya Ulitsa, Moscow 101000, Russia, 
and NRC ``Kurchatov institute'', Moscow, Russia;
e-mail: zabrodin@itep.ru}}

\date{January 2024}
\maketitle

\vspace{-7cm} \centerline{ \hfill ITEP-TH-02/24}\vspace{7cm}

\begin{abstract}

We study the dispersionless limit of the recently introduced Toda lattice
hierarchy with constraint of type B (the B-Toda hierarchy) 
and compare it with that of the DKP and C-Toda hierarchies. 
The dispersionless limits of the B-Toda and C-Toda hierarchies 
turn out to be the same.

\end{abstract}

\end{titlepage}

\vspace{5mm}

%

\tableofcontents

\vspace{5mm}

\section{Introduction}

The Toda hierarchies with constraints of types B and C (or simply 
B- and C-Toda hierarchies) were introduced in the papers \cite{KZ22}
and \cite{KZ21a} respectively as subhierarchies of the 2D Toda lattice
\cite{UT84} defined by imposing certain constraints on the Lax
operators of the latter.
To avoid a confusion, we should stress that the B- and C-Toda hierarchies
introduced in \cite{KZ22,KZ21a} and discussed in this paper
are very different from what was called B- and 
C-Toda in \cite{UT84}. 

The universal dependent variable of the B-Toda hierarchy is the 
tau-function $\tau_n({\bf t})$ which depends on the infinite
set of ``times'' ${\bf t}=\{ t_1, t_2, t_3, \ldots \}$ which are 
in general complex numbers and the discrete variable $n\in \ZZ$. 
As was shown in \cite{PZ23}, the tau-function satisfies an integral
bilinear relation which generates infinite number of bilinear equations
for the tau-function of the Hirota type. Remarkably, this bilinear
relation coincides with that for the tau-function of the so-called
large BKP hierarchy introduced by Jimbo and Miwa as early as in 1983
\cite{JM83} and further studied in \cite{KvdL98,LO15}, see also the book
\cite{book}. The large BKP hierarchy contains a subhierarchy which
connects the tau-functions $\tau_n({\bf t})$ with only even (or odd)
$n$. 
It was rediscovered several times and came to be
known as the DKP hierarchy, the coupled KP hierarchy \cite{HO} and
the Pfaff lattice \cite{AHM,ASM}, see also
\cite{Kakei,IWS,Willox,Takasaki07,Takasaki09}. 
The latter name is motivated by the fact 
that some solutions to the hierarchy are expressed through Pfaffians. 
The solutions and the algebraic structure were studied in 
\cite{Kodama,Kodama1,AKM}, the relation to matrix integrals 
was elaborated in \cite{AHM,ASM,Kakei,Vandeleur,Orlov}.
Bearing certain similarities with the KP and Toda chain hierarchies,
the DKP one is essentially different. 

The dispersionless limit of hierarchies of soliton equations 
consists in re-defining the times as $t_k \to t_k/\hbar$, 
$n\to t_0/\hbar$, rewriting the equations for the tau-function
in terms of $F_{\hbar}=\hbar^2 \log \tau_{t_0/\hbar}({\bf t}/\hbar)$
and setting $\hbar =0$ at the end. This results in an infinite
set of nonlinear partial differential equations for
$$
F(t_0, {\bf t})
=\lim_{\hbar \to 0}\Bigl (\hbar^2 \log \tau_{t_0/\hbar}({\bf t}/\hbar)
\Bigr ).
$$
The dispersionless limit means essentially neglecting higher derivatives
in soliton equations. There is a big literature on different aspects 
of the dispersionless limit, see, e.g., the review \cite{TT95}.

The dispersionless limit of the DKP (dDKP) hierarchy and one-variable
reductions were studied in \cite{AZ14}. The consideration was based on the elliptic parametrization of the dDKP hierarchy which consists in a proper
uniformization of the elliptic curve which is built in the algebraic
structure of the hierarchy. The elliptic modulus of this curve depends
on the times and is one of the dependent variables. 

This paper is devoted to the dispersionless limit of the B-Toda
($=$ large BKP) hierarchy. Along with bilinear relations 
connecting tau-functions with even (odd) indices which are
of the DKP type, this hierarchy contains another equation which 
connects the tau-functions with even and odd indices. We will show
that this additional equation (in the dispersionless limit) makes
the elliptic curve degenerate (rational), and the dispersionless 
limit of the B-Toda turns out to be the same as that of the C-Toda
elaborated in \cite{TZ22}. In fact this result might be expected from
consideration of the constraints for the Lax operators (it is easy to
see that their dispersionless limit is the same in 
the B- and C-cases) but the analysis
on the level of tau-functions seems to be instructive.

\section{Bilinear equations of the B-Toda hierarchy}

In order to write down the integral bilinear relation for the 
tau-function $\tau_n({\bf t})$ of the B-Toda hierarchy obtained in
\cite{PZ23}, we prepare
the standard notation:
\beq\label{b1}
\xi ({\bf t}, z)=\sum_{k\geq 1}t_k z^k,
\eeq
\beq\label{b2}
{\bf t}\pm [z^{-1}]=\Bigl \{ t_1\pm z^{-1}, t_2\pm \frac{1}{2}\, z^{-2},
t_3\pm \frac{1}{3}\, z^{-3}, \, \ldots \Bigr \}.
\eeq
The equation reads:
\beq\label{b3}
\begin{array}{l}
\displaystyle{
\frac{1}{\pi i}
\oint_{C_{\infty}} \Bigl [z^{n-n'-2}e^{\xi ({\bf t}-{\bf t}',z)}
\tau_{n-1}({\bf t}-[z^{-1}])\tau_{n'+1}({\bf t}'+[z^{-1}])}
\\ \\
\displaystyle{\phantom{aaaaaaaa}
+\, z^{n'-n-2}e^{-\xi ({\bf t}-{\bf t}',z)}
\tau_{n+1}({\bf t}+[z^{-1}])\tau_{n'-1}({\bf t}'-[z^{-1}])\Bigr ]dz}
\\ \\
\displaystyle{\phantom{aaaaaaaaaaaaaaaaaaaaaaaaaaaaaaaaaa}
=(1\! -\! (-1)^{n-n'})
\tau_{n}({\bf t})\tau_{n'}({\bf t}').}
\end{array}
\eeq
It is valid for all $n, n'\in \ZZ$ and ${\bf t}$, ${\bf t}'$. 
Here $C_{\infty}$ is a big contour around $\infty$ (with positive 
orientation). Although the origin of the two hierarchies is different,
the same equation holds for the tau-function of the 
large BKP hierarchy (see, e.g., 
\cite{book}).
Putting $n-n'$, ${\bf t}-{\bf t}'$ to some special values, one can obtain
from (\ref{b3}) 
bilinear equations of the Hirota-Miwa type. 

First, put $n'=n$, ${\bf t}-{\bf t}'=[a^{-1}]+[b^{-1}]$. Then
$$
e^{\xi ({\bf t}-{\bf t}',z)}=\frac{ab}{(a-z)(b-z)}
$$
and the residue calculus yields:
\beq\label{b4}
\begin{array}{l}
\displaystyle{
a^2 \frac{\tau_{n+1}({\bf t}+[b^{-1}])
\tau_{n-1}({\bf t}+[a^{-1}])}{\tau_{n+1}({\bf t}+[a^{-1}]+[b^{-1}])
\tau_{n-1}({\bf t})}-
b^2 \frac{\tau_{n+1}({\bf t}+[a^{-1}])
\tau_{n-1}({\bf t}+[b^{-1}])}{\tau_{n+1}({\bf t}+[a^{-1}]+[b^{-1}])
\tau_{n-1}({\bf t})}}
\\ \\
\phantom{aaaaaaaaaaa}
\displaystyle{
=\, a^2-b^2 -(a-b)\p_{t_1}
\log \frac{\tau_{n+1}({\bf t}+[a^{-1}]+[b^{-1}])}{\tau_{n-1}({\bf t})}}.
\end{array}
\eeq

Second, differentiate equation (\ref{b3}) with respect to $t_1$
and put $n-n'=2$, ${\bf t}-{\bf t}'=[a^{-1}]+[b^{-1}]$.
The residue calculus yields:
\beq\label{b5}
\begin{array}{l}
\displaystyle{
\frac{\tau_{n}({\bf t})
\tau_{n}({\bf t}+[a^{-1}]+[b^{-1}])}{\tau_{n}({\bf t}+[a^{-1}])
\tau_{n}({\bf t}+[b^{-1}])}-
\frac{\tau_{n-2}({\bf t})
\tau_{n+2}({\bf t}+[a^{-1}]+[b^{-1}])}{a^2b^2\tau_{n}({\bf t}+[a^{-1}])
\tau_{n}({\bf t}+[b^{-1}])}}
\\ \\
\phantom{aaaaaaaaaaa}
\displaystyle{
=\, 1-\frac{1}{a-b}\, \p_{t_1}
\log \frac{\tau_{n}({\bf t}+[a^{-1}])}{\tau_{n}({\bf t}+[b^{-1}])}}.
\end{array}
\eeq
The two equations (\ref{b4}), (\ref{b5}) are equations of the 
DKP hierarchy \cite{Takasaki07}. They connect tau-functions with
indices of the same parity (even or odd).

Another equation connects tau-functions with even $n$ with tau-functions
with odd $n$. It is obtained from (\ref{b3}) at 
$n-n'=1$, ${\bf t}-{\bf t}'=[a^{-1}]+[b^{-1}]$.
The residue calculus yields:
\beq\label{b6}
\begin{array}{l}
\displaystyle{
a\frac{\tau_{n}({\bf t}+[a^{-1}])
\tau_{n+1}({\bf t}+[b^{-1}])}{\tau_{n}({\bf t})
\tau_{n+1}({\bf t}+[a^{-1}]+[b^{-1}])}-
b\frac{\tau_{n}({\bf t}+[b^{-1}])
\tau_{n+1}({\bf t}+[a^{-1}])}{\tau_{n}({\bf t})
\tau_{n+1}({\bf t}+[a^{-1}]+[b^{-1}])}}
\\ \\
\phantom{aaaaaaaaaaa}
\displaystyle{
=\, a-b+(a^{-1}\! -\! b^{-1})\, 
\frac{\tau_{n-1}({\bf t})
\tau_{n+2}({\bf t}+[a^{-1}]+[b^{-1}])}{\tau_{n}({\bf t})
\tau_{n+1}({\bf t}+[a^{-1}]+[b^{-1}])}}.
\end{array}
\eeq

\section{The dispersionless limit}

In the dispersionless limit one should re-scale the independent
variables as
$t_k \to t_k/\hbar$, $n\to t_0/\hbar$ and consider solutions (tau-functions)
that have an essential singularity at $\hbar =0$ and have the form
\beq\label{d1}
\tau_{t_0/\hbar}({\bf t}/\hbar )=e^{\frac{1}{\hbar^2}F(t_0, {\bf t}, \hbar )}
\eeq
as $\hbar \to 0$, where $F$ is a smooth function of $t_0$ and ${\bf t}$
having a regular expansion in $\hbar$ as $\hbar \to 0$. The function
$F=F(t_0, {\bf t}, 0)$ is sometimes called the dispersionless tau-function
(although the $\hbar \to 0$ limit of the tau-function itself does not exist).

Let us introduce the differential operator
\beq\label{D(z)}
D(z)=\sum_{k\geq 1}\frac{z^{-k}}{k}\, \p_{t_k}.
\eeq
For small but non-zero $\hbar$ equations (\ref{b4}), (\ref{b5}), 
(\ref{b6}) can be identically rewritten in terms of the function 
$F$ in the following form. Equation (\ref{b4}):
\beq\label{d2}
\begin{array}{l}
\displaystyle{
a^2 \exp \left (\frac{1}{\hbar^2}\Bigl (e^{\hbar \p_{t_0} +\hbar D(b)}F+
e^{-\hbar \p_{t_0} +\hbar D(a)}F- e^{-\hbar \p_{t_0}}F - 
e^{\hbar \p_{t_0} +\hbar D(a)+\hbar D(b)}F\Bigr )\right )}
\\ \\
\displaystyle{
-\, b^2 \exp \left (\frac{1}{\hbar^2}\Bigl (e^{\hbar \p_{t_0} +\hbar D(a)}F+
e^{-\hbar \p_{t_0} +\hbar D(b)}F- e^{-\hbar \p_{t_0}}F - 
e^{\hbar \p_{t_0} +\hbar D(a)+\hbar D(b)}F\Bigr )\right )}
\\ \\
\phantom{aaaaaaaaaaa}
=\, \displaystyle{
a^2-b^2 -\hbar^{-1} (a-b)\p_{t_1} \Bigl ( 
e^{\hbar \p_{t_0} +\hbar D(a)+\hbar 
D(b)}F- e^{-\hbar \p_{t_0}}F\Bigr )}.
\end{array}
\eeq
Equation (\ref{b5}):
\beq\label{d3}
\begin{array}{l}
\displaystyle{
\exp \left (\frac{1}{\hbar^2}\Bigl (F+
e^{\hbar D(a)+\hbar D(b)}F- e^{\hbar D(a)}F - 
e^{\hbar D(b)}F\Bigr )\right )}
\\ \\
\displaystyle{
-\, \frac{1}{a^2b^2} \exp \left (\frac{1}{\hbar^2}\Bigl (
e^{-2\hbar \p_{t_0}}F+
e^{2\hbar \p_{t_0} +\hbar D(a)+\hbar D(b)}F- e^{\hbar D(a)}F - 
e^{\hbar D(b)}F\Bigr )\right )}
\\ \\
\phantom{aaaaaaaaaaa}
=\, \displaystyle{
1-\frac{\hbar^{-1}}{a-b}\, \p_{t_1} \Bigl ( 
e^{\hbar D(a)}F-
e^{\hbar D(b)}F\Bigr )}.
\end{array}
\eeq
Equation (\ref{b6}):
\beq\label{d4}
\begin{array}{l}
\displaystyle{
a\exp \left (\frac{1}{\hbar^2}\Bigl (e^{\hbar D(a)}F+
e^{\hbar \p_{t_0} +\hbar D(b)}F-F - 
e^{\hbar \p_{t_0} +\hbar D(a)+\hbar D(b)}F\Bigr )\right )}
\\ \\
\displaystyle{
-\, b\exp \left (\frac{1}{\hbar^2}\Bigl (e^{\hbar D(b)}F+
e^{\hbar \p_{t_0} +\hbar D(a)}F-F - 
e^{\hbar \p_{t_0} +\hbar D(a)+\hbar D(b)}F\Bigr )\right )-(a-b)}
\\ \\
=\, \displaystyle{
(a^{-1}\! -\! b^{-1})\exp \left (\frac{1}{\hbar^2}
\Bigl (e^{-\hbar \p_{t_0}}F+
e^{2\hbar \p_{t_0} +\hbar D(a)+\hbar D(b)}F-F - 
e^{\hbar \p_{t_0} +\hbar D(a)+\hbar D(b)}F\Bigr )\right )
}.
\end{array}
\eeq

In this form, the limit $\hbar \to 0$ is straightforward.
We obtain the following equations. Equation (\ref{d2}) gives:
\beq\label{d5}
e^{-D(a)D(b)F}\,
\frac{a^2 e^{-2\p_{t_0}D(a)F}-b^2 e^{-2\p_{t_0}D(b)F}}{a-b}
=a+b -\p_{t_1}\Bigl (2\p_{t_0}+D(a)+D(b)\Bigr )F.
\eeq
Equation (\ref{d3}) gives:
\beq\label{d6}
1-\frac{\p_{t_1}D(a)F -\p_{t_1}D(b)F}{a-b}=
e^{D(a)D(b)F} -\frac{1}{a^2b^2}\, e^{(2\p_{t_0}+D(a))(2\p_{t_0}+D(b))F}.
\eeq
These two are equations of the dDKP hierarchy. In the dispersionless
B-Toda (or large BKP) hierarchy they are supplemented by additional 
equation which is the $\hbar \to 0$ limit of (\ref{d4}):
\beq\label{d7}
e^{-D(a)D(b)F}\, \Bigl (ae^{-\p_{t_0}D(a)F}-be^{-\p_{t_0}D(b)F}\Bigr )
=(a-b)\Bigl (1-\frac{1}{ab}\, e^{\p_{t_0}(2\p_{t_0}+D(a)+D(b))F}\Bigr ).
\eeq
These equations, being expanded in negative powers of $a$, $b$, 
generate an infinite number of nonlinear partial differential equations
for the function $F$. For example, the simplest equation contained
in (\ref{d7}) reads
\beq\label{d8}
F_{02}-2F_{11}-F_{01}^2 +2e^{2F_{00}}=0,
\eeq
where we denote $F_{mn}=:\p_{t_m}\p_{t_n}F$. 

We will show that the latter equation, (\ref{d7}), 
makes the first two equations (\ref{d5}),
(\ref{d6}) equivalent, and each of them, in its turn, is equivalent
to (\ref{d7}). To this end, let us rewrite these equations in a 
more suggestive form. Introduce the functions
\beq\label{d9}
p(z)=z-\p_{t_1}\nabla (z)F, \quad w(z)=ze^{-\p_{t_0}\nabla (z)F},
\eeq
where $\nabla (z)$ is the differential operator
\beq\label{d10}
\nabla (z)=\p_{t_0}+D(z).
\eeq
In terms of these functions equations (\ref{d5})--(\ref{d7}) acquire
the following form. Equation (\ref{d5}):
\beq\label{d11}
e^{-\nabla (a)\nabla (b)F+F_{00}}\,
\frac{w^2(a)-w^2(b)}{w(a)w(b)}=(b^{-1}-a^{-1})(p(a)+p(b)).
\eeq
Equation (\ref{d6}):
\beq\label{d12}
e^{\nabla (a)\nabla (b)F+F_{00}}\, \frac{w^2(a)w^2(b)-1}{w(a)w(b)}=
\frac{p(a)-p(b)}{b^{-1}-a^{-1}}.
\eeq
Equation (\ref{d7}):
\beq\label{d13}
(b^{-1}-a^{-1})e^{\nabla (a)\nabla (b)F}=
\frac{w(a)-w(b)}{w(a)w(b)-1}.
\eeq
Tending $b\to \infty$ in equation (\ref{d13}), we obtain, in the
order $O(b^{-1})$:
$$
a-\p_{t_1}\nabla (a)F=e^{F_{00}}\Bigl (w(a)-w^{-1}(a)\Bigr ),
$$
or
\beq\label{d14}
p(a)=R\Bigl (w(a)-w^{-1}(a)\Bigr ), \quad R=:e^{F_{00}}.
\eeq
Multiplying the two equations (\ref{d11}), (\ref{d12}), we obtain
the following relation:
$$
p^2(a)-R^2 \Bigl (w^2(a)+w^{-2}(a)\Bigr )=
p^2(b)-R^2 \Bigl (w^2(b)+w^{-2}(b)\Bigr )
$$
which means that the combination
$p^2(z)-R^2 \Bigl (w^2(z)+w^{-2}(z)\Bigr )$ does not depend on $z$.
Tending $z\to \infty$, we get the equation connecting the
functions $p(z)$, $w(z)$ and defining an elliptic curve:
\beq\label{d15}
p^2(z)=R^2 \Bigl (w^2(z)+w^{-2}(z)\Bigr )+V, \quad 
V=F_{02}-2F_{11}-F_{01}^2.
\eeq
This is the elliptic curve involved in the dispersionless DKP hierarchy
noticed in \cite{Takasaki07} and elaborated in \cite{AZ14}. 
However, in our case, when we have the additional equation 
(\ref{d7}) and its corollary (\ref{d8}) which states that $V=-2R^2$, 
the elliptic curve degenerates and becomes rational:
$$
p^2(z)=R^2 \Bigl (w(z)-w^{-1}(z)\Bigr )^2,
$$
which is simply (\ref{d14}) after taking the square roots of the both 
sides.
Finally, after plugging (\ref{d14}) into equation (\ref{d12}), 
one can see that it becomes equivalent to (\ref{d13}). 

\section{Conclusion}

We have considered the dispersionless limit of the Toda hierarchy
with constraint of type B (the B-Toda hierarchy) on the level
of tau-function. The dispersionless 
version of the hierarchy turns out to be the same as that of the 
C-Toda hierarchy studied previously in \cite{TZ22}. This fact
could be expected from the form of the constraints for the Lax
operators for the two hierarchies: in the dispersionless limit
they coincide, like for the BKP and CKP hierarchies. 

We have also compared our result with the dispersionless version
of the DKP (Pfaff lattice) hierarchy. The latter is a subhierarchy 
of the B-Toda consisting of bilinear equations for tau-functions
with indices of the same parity. In the dispersionless limit, 
these equations lead to an elliptic curve built in the structure
of the hierarchy. In the B-Toda case, there are additional bilinear
equations connecting tau-functions with indices of different parity.
We have shown that in this case the elliptic curve degenerates
and becomes rational. 

The geometric meaning of the dispersionless limit of the B-Toda
hierarchy, as soon as it coincides with that of the C-Toda,
is clear from the results of the paper \cite{TZ22}. Equations 
of the dispersionless B-Toda hierarchy govern conformal maps
of planar domains symmetric with respect to the real axis
(reflection-symmetric domains). The independent variables 
of the hierarchy are identified with the (real) harmonic 
moments of such domains. One-variable reductions of the hierarchy
are described by solutions to an ordinary differential equation 
of L\"owner type which was called the symmetric L\"owner equation
in \cite{TZ22}. The equation contains an arbitrary function
(the ``driving function'') which characterizes the reduction. 
Geometrically, solutions of the symmetric L\"owner equation 
are conformal mapping functions which send the unit circle with
two symmetric slits to the unit circle, and the driving function
parametrizes the shape of the slits.

\section*{Acknowledgments}

\addcontentsline{toc}{section}{Acknowledgments}

The author thanks V. Prokofev for discussions. 
This work is an output of a research project 
implemented as a part of the Basic Research Program 
at the National Research University Higher School 
of Economics (HSE University).

\end{document}